\documentclass[]{spie}  
\usepackage[]{graphicx}

\title{Linear and Non-linear Response of Lithographically Defined Plasmonic Nanoantennas} 

\author{K. Schraml\supit{a}, M. Kaniber\supit{a}, J. Bartl\supit{a}, G. Glashagen\supit{a}, A. Regler\supit{a}, T.Campbell\supit{b} and J.J. Finley\supit{a}
\skiplinehalf
\supit{a}Walter Schottky Institut and Physik Department, Technische Universit\"at M\"unchen,\\ Am Coulombwall 4, 85748 Garching, Germany;\\
\supit{b}Work performed while at the Institute for Critical Technology and Applied Science (ICTAS), Virginia Polytechnic Institute and State University (Virginia Tech), Blackburg, VA 24061, USA}

\authorinfo{Corresponding author: michael.kaniber@wsi.tum.de,  www.wsi.tum.de}

\begin{document} 
  \maketitle 

\begin{abstract}

We present numerical studies, nano-fabrication and optical characterization of bowtie nanoantennas demonstrating their superior performance with respect to the electric field enhancement as compared to other Au nanoparticle shapes. For optimized parameters, we found mean intensity enhancement factors \textgreater$2300\times$ in the feed-gap of the antenna, decreasing to 1300\,$\mathrm{\times}$ when introducing a 5\,nm titanium adhesion layer. Using electron beam lithography we fabricated gold bowties on various substrates with feed-gaps and tip radii as small as 10\,nm. In polarization resolved measurement we experimentally observed a blue shift of the surface plasmon resonance from 1.72\,eV to 1.35\,eV combined with a strong modification of the electric field enhancement in the feed-gap. Under excitation with a 100\,fs pulsed laser source, we observed non-linear light emission arising from two-photon photoluminescence and second harmonic generation from the gold. The bowtie nanoantenna shows a high potential for outstanding conversion efficiencies and the enhancement of other optical effects which could be exploited in future nanophotonic devices.  
\end{abstract}


\keywords{surface plasmon, nanoantenna, bowtie, non-linear optics, electric field enhancement }

\section{INTRODUCTION}
\label{sec:intro}  

Plasmonic nanostructures provide the potential to focus and localize electromagnetic fields to extreme sub-wavelength dimension and, thereby, dramatically enhance the strength of linear light matter interactions and non-linear optical phenomena\cite{stockman2004nanofocusing,schuller2010plasmonics,ciraci2012probing,ozbay2006plasmonics,gramotnev2010plasmonics}. The strong localization of light obviates the need for phase matching significantly enhancing the strength of nonlinear processes and, therefore, potentially opens the way towards all optical switching devices\cite{kauranen2012nonlinear}. In the past years, various plasmonic nanostructures have been investigated with respect to their ability to confine light into deep subwavelength dimensions. The surface plasmon resonance and the electric fields were analyzed as a function of particle size\cite{muhlschlegel2005resonant,fischer2008engineering,hanke2009efficient,prangsma2012electrically}, shape\cite{fischer2008engineering,sonnichsen2002drastic} and feed-gap size\cite{hanke2009efficient,fromm2004gap,schuck2005improving,merlein2008nanomechanical}, wavelength range\cite{fromm2004gap,rivas2004propagation,hibbins2005experimental} and materials\cite{wang2006general}.

In this paper, we present numerical and experimental studies of different plasmonic nanostructures. We identify the bowtie nanoantenna as an advantageous geometry that provides electric field intensity enhancement factors $FE \geq 4500$  that can be deterministically controlled with nanometer precision. Similar simulations were employed to identify an optimal gold film thickness of $t\,=\,35$\,nm and to test the influence of a 5\,nm thick titanium adhesion layer on the plasmonic properties. Using electron beam lithography we realized high quality bowtie resonators with feed-gaps and tip radii as small as 10\,nm on both, insulating and semiconducting substrates. Reflectivity measurements revealed a strong dependence of the surface plasmon resonance on the polarization of the incident light since we observe a blue shift from 1.72\,eV to 1.35\,eV when rotating the polarization by $90^\circ$ with respect to the long bowtie axis. According to our simulations, the electric field enhancement in the feed-gap completely vanishes under these conditions, which is also revealed in our optical experiments. When exciting with a 100\,fs pulsed laser source, we observe two-photon photoluminescence\cite{PhysRevB.68.115433,biagioni2009dependence,bouhelier2005surface,ko2010nonlinear,schuck2005improving} and second harmonic generation\cite{hubert2007role,mcmahon2006second,ko2010nonlinear,hanke2009efficient}, which can be completely switched off by rotating a polarizer in the excitation path. Furthermore, we did not observe any non-linear signal when focusing our laser at identical excitation power on an unpatterned gold film indicating that the non-linear processes occur only due to the presence of the electric field enhancement in the feed-gap of the nanoantenna.

\section{Optimization of Plasmonic Nanoantennas for Non-Linear Optical Processes} 
\label{sec:optimization}

To identify which antenna geometry produces the highest electric field enhancement and a surface plasmon resonance in the visible to near infrared spectral range, we performed finite difference time domain (FDTD) simulations using a commercially available software package\cite{test}. We calculate the surface plasmon resonance and the electric field enhancement using  a total field scattered field (TFSF) source polarized along the long bowtie axis (x-axis), perfectly matched layers as boundary conditions and a mesh size of 1\,nm in the vicinity of the investigated gold structures. Throughout our studies we use gold\cite{johnson1972optical} since it is chemically stable and, therefore, allows reproducible experiments. We varied the shape and thickness of the particles and tested the impact of adding a titanium adhesion layer. The results of this studies are presented in the following.

\subsection{Geometry of Plasmonic Antenna} 
\label{sec:geometry}

   \begin{figure}[b]
   \begin{center}
   \includegraphics[width=\columnwidth]{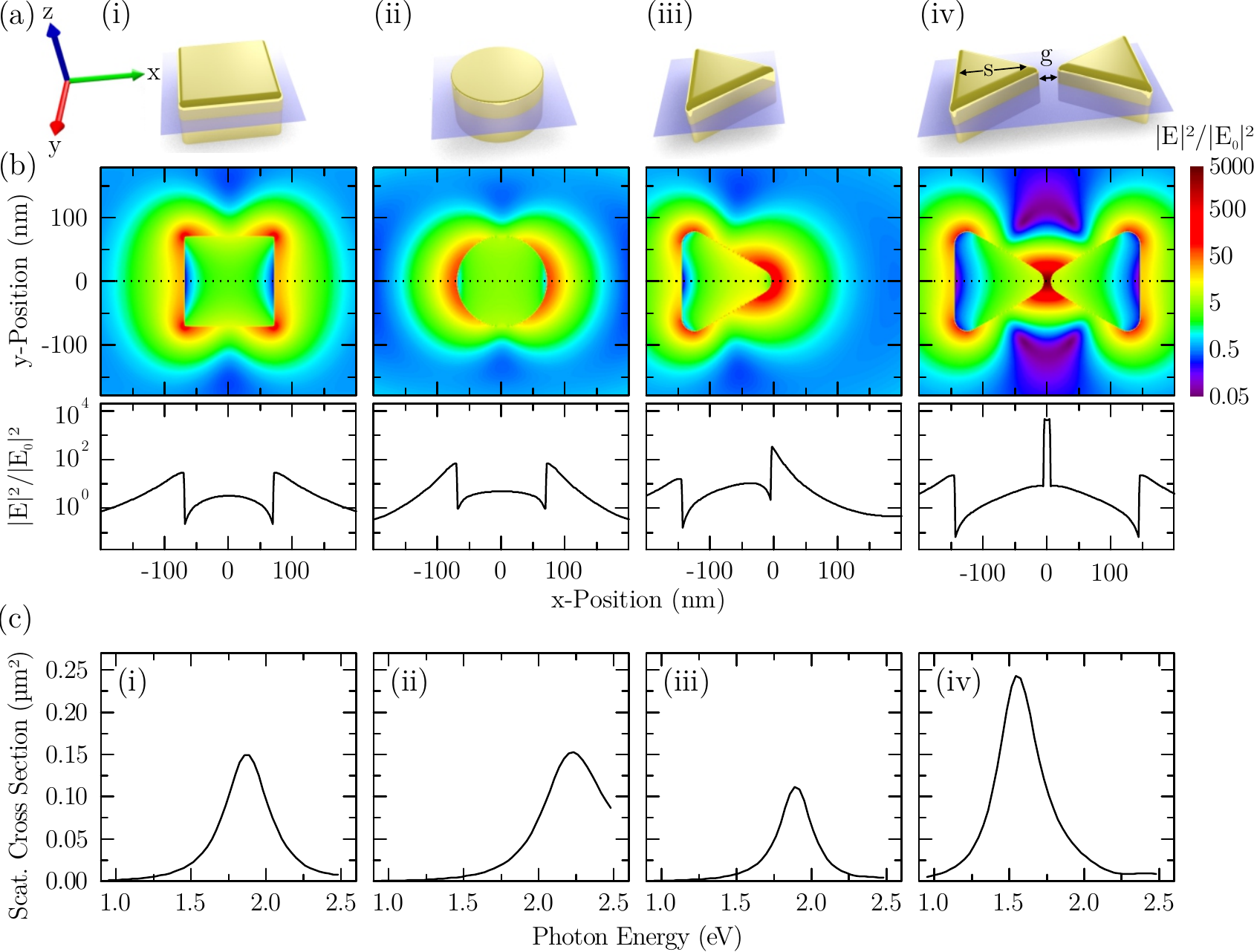}
   \end{center}
	 \caption{\label{fig:geometry}FDTD simulations of lithographically defined plasmonic nanoparticles. (a) Sketch of the investigated cubic (i), cylindrical (ii), triangular (iii) and bowtie (iv) nanoantenna. (b) Upper panels show 2D-plot of the electric field enhancement along the blue plane indicated in (a). Lower panels show the corresponding cross sections along the dotted lines in the upper panels. (c) Calculated Scattering cross section of the investigated structures.}
   \end{figure} 

In a first series of simulations we investigated different particle geometries to identify the shape which provides the largest electric field enhancements. As sketched in Fig. \ref{fig:geometry}(a), we studied a cubic (i), cylindrical (ii), and triangular (iii) single particle and a bowtie nanoantenna (iv) consisting of two triangles in a tip-to-tip arrangement. In all simulations, the particles have a thickness $t=35$\,nm and a size $s=140$\,nm which is defined as the length along the x-axis. Fig. \ref{fig:geometry}(b) shows the corresponding electric field intensity enhancement, which is defined as the ratio of the steady state electric field intensity $|E|^2$ compared to the electric field of the incoming plane wave $|E_0|^2$. We plot the cross section along the x-y plane through the center of the particles as indicated by the blue planes in (a) on a logarithmic scale. To facilitate a direct comparison, we also plotted the cross section along the dotted lines in the lower part of the panel bisecting the midpoint of various particle geometries. At the gold/air interface at the edge of of the particles along the x-axis, we find a maximum enhancement factor of $FE_{\diamond}=28$ for the cube, $FE_{\circ}=70$ for the cylinder and $FE_{\triangleleft}=340$ for the triangle, respectively. The strong increase for the triangle is due to the well known lightening rod effect which leads to a strong enhancement of the electric field near tips and corners. Further improvement can be achieved by adding a second triangle in a tip-to-tip configuration to the system. The lightening rod effect of the second particle and, most importantly, the near-field coupling between the two triangles boosts the electric field enhancement in its feed-gap $g$ by more than one order of magnitude to $FE_{ \triangleright \triangleleft} \geq 4500$. This so-called bowtie nanoantenna\cite{novotny2011antennas} is, therefore, very well suited to generate extremely high electric field enhancements which can be used to enhance non-linear optical effects and other nanophotonic processes. 

In a second step we calculated the scattering cross sections $\sigma$ of the investigated nanostructures as a function of the incident photon energies. As depicted in Fig. \ref{fig:geometry}(c), we observe a clear resonance which can be attributed to the electric dipole mode of the localized surface plasmon polaritons. For the single particles the peak energy $E_{SPR}$ is altered despite the fact that all particles have the same geometrical length along the x-axis. For the cube and the triangle we calculate a peak in the scattering cross section for $E_{SPR,\diamond}=1.87$\,eV and $E_{SPR,\triangleright}=1.89$\,eV, respectively, whereas the cylinder peaks at $E_{SPR,\circ}=2.23$\,eV. Furthermore, we observe a significant difference in the linewidth (full width at half maximum) $\Delta E_{SPR}$ of the resonances. For the cylinder we found a value of $\Delta E_{SPR,\circ}=0.53$\,eV, compared to $\Delta E_{SPR,\diamond}=0.34$\,eV for the cube, a difference that originates from increased damping due to interband transitions in the gold, which sets in for photon energies $E_{ph}\geq 1.8$\,eV and increases with energy\cite{PhysRevB.68.115433}. For the triangle we observe a value of $\Delta E_{SPR,\triangleright}=0.25$\,eV which can be explained by decreased radiation damping due to the lower volume compared to the other particles\cite{wokaun1982radiation}. In contrast, the bowtie shows again an increased scattering cross section that obviously originates from an increased geometrical cross section. Furthermore, the interparticle coupling redshifts and broadens the resonance to $E_{SPR,\triangleright \triangleleft}=1.55$\,eV and $\Delta E_{SPR,\triangleright \triangleleft}=0.327$\,eV, respectively\cite{schraml2014optical}. 

In summary, our simulations illustrate how the peak energy and the width of the surface plasmon resonance strongly depend on the shape of the particle as already discussed in the literature\cite{fischer2008engineering,sonnichsen2002drastic}. The bowtie-geometry provides a high scattering cross section and a moderate damping. In combination with the high electric field enhancement, it is, therefore, a superior particle geometry to investigate linear and non-linear optical phenomena.

			\subsection{Gold Layer Thickness Dependence} 
			\label{subsec:thickness}
			
   \begin{figure}[ht]
   \begin{center}
   \includegraphics[width=\columnwidth]{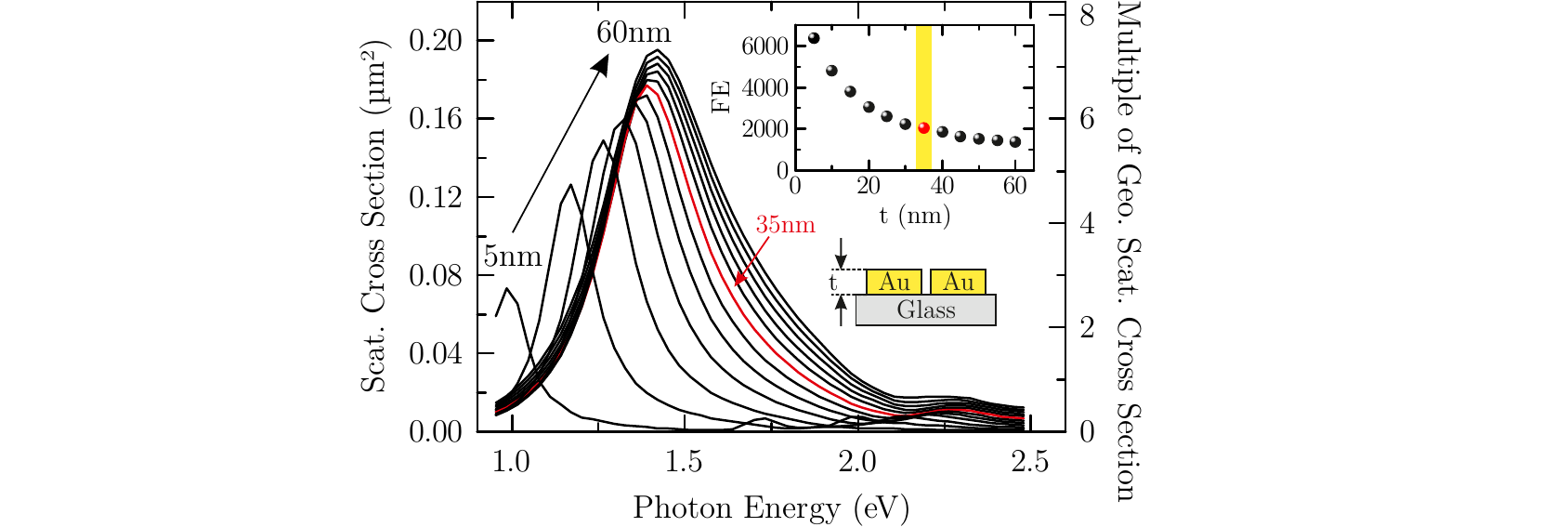}
   \end{center}
   \caption{\label{fig:thickness} Simulated scattering cross sections of bowties on a glass substrate for various metal film thicknesses t. (Inset) Electric field enhancement FE as a function of bowtie thickness.}
      \end{figure} 

In addition to the geometry in the x-y plane, the peak energy of the surface plasmon resonance is also strongly influenced by the thickness t of the gold film in z-direction. Therefore, we simulated the scattering cross section and the electric field enhancement as a function of this parameter. The obtained results are plotted in Fig. \ref{fig:thickness}; for a $t=5$\,nm thick gold layer, the surface plasmon resonance peaks at $E_{SPR,5}=0.99$\,eV and red shifts with increasing thickness. At around $t=30-40$\,nm the peak energy saturates at $E_{SPR}\sim 1.42$\,eV. Simultaneously, the maximum of the scattering cross section increases from $\sigma = 0.74\,\mathrm\mu m^2$ and saturates at $\sigma \sim 0.20\,\mu m^2$, almost 8 times larger than the geometrical scattering cross section as plotted for comparison on the right y-axis of Fig. \ref{fig:thickness}. The inset shows the mean value of the calculated electric field intensity enhancement inside the feed gap at the air/glass interface. We observe a decrease from $FE=6350$ to $FE\sim 1350$ with increasing thickness. Consequently, the structures should be as thin as possible, in order to obtain the highest enhancement values. However, since we want to operate in the visible to near-infrared spectral range, we choose a value of $t=35$\,nm as a good trade-off between field enhancement and a suitable frequency for the surface plasmon resonance.

			\subsection{Influence of a Titanium Adhesion Layer on Plasmonic Properties} 
			\label{subsec:adhesion}
   \begin{figure}[hb]
   \begin{center}
   \includegraphics[width=\columnwidth]{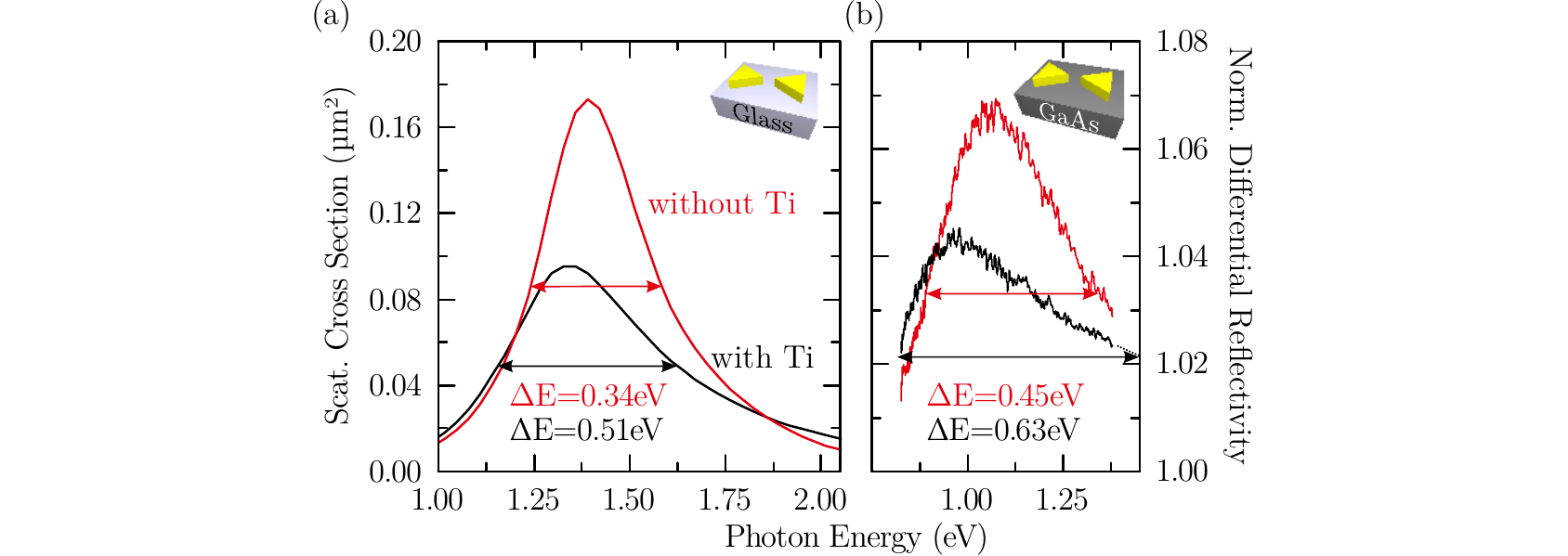}
   \end{center}
   \caption{\label{fig:adhesion}Simulated scattering cross sections for bowties ($s=140$\,nm, $g=10$\,nm) on a glass substrate with (black) and without (red) a 5\,nm titanium adhesion layer. (b) Corresponding experimental differential reflectivity spectra for a bowtie on a GaAs substrate ($s=170$\,nm, $g=10$\,nm) with (black) and without (red) a 5\,nm titanium adhesion layer.}
   
   \end{figure} 

Until now we simulated only gold structures defined directly on a glass substrate. In reality the adhesion of gold on glass is poor\cite{benjamin1960adhesion,goss1991application} and adhesion layers, usually thin chromium  or titanium layers, are widely used. For this reason, we implemented a 5\,nm thick titanium layer in our simulation scheme and tested its influence on the plasmonic properties of the bowtie. The spectrum of the scattering cross section is plotted in Fig. \ref{fig:adhesion}(a) with and without the Ti layer between the gold and a glass substrate. We observe that the peak value drops by 45\% from $\sigma_{SiO_2} =0.173\,\mu m^2$ to $\sigma_{Ti} =0.096\,\mu m^2$, while red shifting from $E_{SPR,SiO_2}=1.39$\,eV to $E_{SPR,Ti}=1.34$\,eV. Simultaneously, the line width broadens from $\Delta E_{SPR,SiO_2} = 0.34$\,eV to $\Delta E_{SPR,Ti}=0.51$\,eV indicating an increased damping which also decreases the mean field enhancement in the feed-gap from $FE_{SiO_2}=2300$ to $FE_{Ti}=1300$. This effect originates from the higher imaginary part of the Ti-dielectric function compared to gold. For a photon energy of $E_{ph}=1.39$\,eV the literature reports a value of $\epsilon_{i,Ti}=26.06$ compared to $\epsilon_{i,Au}=1.93$\cite{johnson1972optical,johnson1974optical}. Fig. \ref{fig:adhesion}(b) shows experimentally determined scattering spectra/normalized differential reflectivity $\gamma$ of bowties on a GaAs wafer ($s=170$\,nm, $g=10$\,nm) with (black) and without (red) a 5\,nm titanium adhesion layer measured as discussed in the next section. As expected the peak value reduces from $\gamma =0.067$ to $\gamma=0.043$ while redshifting the resonance from $E_{SPR,exp}=1.06$\,eV to $E_{SPR,exp,Ti}=0.97$\,eV. Therefore, we clearly could reproduce the simulated behavior experimentally proving the high accuracy and reliability of the applied numerical simulations. We note that the offset in photon energies and the increased linewidth is related to the higher refractive index of GaAs compared to glass and the difference in size as recently discussed\cite{schraml2014optical}. From our findings we conclude that in non-linear optics experiments an adhesion layer should be avoided whenever possible to ensure the lowest damping and highest field enhancements.

\section{Fabrication \& Optical Studies} 
\label{sec:fabrication}

   \begin{figure}[b]
   \begin{center}
   \includegraphics[width=\columnwidth]{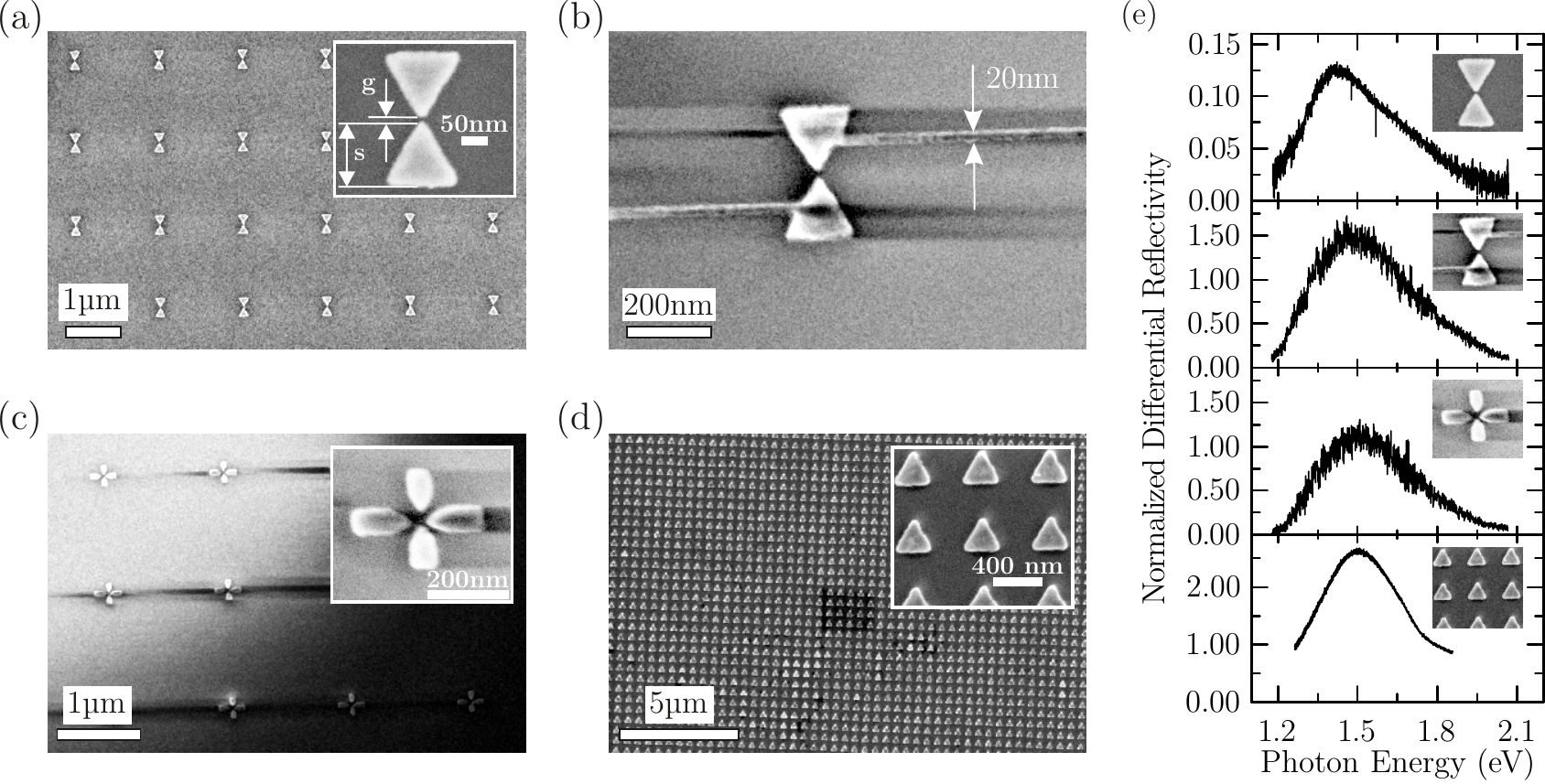}
   \end{center}
   \caption{\label{fig:fabrication}SEM images of various gold nanostructures on different substrates. (a) Bowtie array on Si/$\mathrm{SiO_2}$. (Inset) Closeup of a single bowtie on GaAs. (b) Electrically connected bowtie on glass. (c) Cross antennas on glass. (d) Large scale triangle array on GaAs. (Inset) Closeup of individual triangles. (e) Measured normalized differential reflectivity spectra for shown structures.}
   
   \end{figure} 

Having determined the optimal parameters via our simulations, we fabricated bowties using standard electron beam lithography and metalization followed by a lift-off process. In a first step, we spin coated a 70\,nm think PMMA 950K electron beam resist onto our substrates. After the bake out, the sample structures were written in a Raith E-line system using an acceleration voltage of 30\,kV. The development in a Methylisobutylketon solution diluted with Isopropanol (1:3) was followed by evaporation of a 35\,nm thin gold film. Finally, the lift-off was performed in warm Acetone revealing our high quality nanostructures. For non-conductive substrates we evaporated a 10\,nm thick aluminum film onto the electron beam resist to avoid charging effects during the illumination with electrons. Before development, this layer was then etched away using a metal-ion-free photoresist developer. Furthermore, it was necessary to evaporate a 5\,nm titanium layer below the gold on the glass substrates to provide sufficient adhesion. Further details on the fabrication can be found elsewhere\cite{schraml2014optical}. 

In Fig. \ref{fig:fabrication} we show a selection of representative collection of scanning electron microscope (SEM) images that were taken from structures made using the fabrication process described above. Panel (a) depicts an array of optimized bowties on a Si/$\mathrm{SiO_2}$ substrate and a close up of a single bowtie on GaAs. The antennas are arranged in a square lattice with a pitch of $1.5\, \mu m$ to avoid near- and far-field coupling between them. With our method we control the size, the feed-gaps and the tip-radii with a excellent precision below 10\,nm on various substrates which is extremely beneficial to achieve high electric field enhancement\cite{schuck2005improving}. Furthermore, we can equip our bowties with 20\,nm thin electrical contacts as shown in panel (b) for a nanoantenna on glass. The flexibility of lithographically defined structures allows the realization of practically any structure on different substrates. To show just one further example, we fabricated cross-antennas on a glass substrate (Fig. \ref{fig:fabrication}(c)) that can be used to engineer the plasmonic response to the polarization\cite{biagioni2009cross} of the incident light field. In addition, large scale structures such as the triangle array shown in Fig. \ref{fig:fabrication} (f) are easily producible with the introduced recipe, as well. In a nutshell, the method provides us with a powerful tool for highly flexible design and fabrication of any plasmonic nanostructure with state of the art precision. 

To probe the surface plasmon resonance of the defined structures, we measured the white light reflectivity spectrum of the nanoantennas $R$ and the substrate $R_0$, respectively, in the wavelength range from 600\,nm to 1000\,nm. By calculating the normalized differential reflectivity $\gamma$, defined by $\gamma = \Delta R/R_0$, where $ \Delta R=R-R_0$, we reveal the spectral form of the surface plasmon resonance\cite{schraml2014optical}. For excitation we use a super white light continuum source\cite{fianium}, whose coherent white light can be focused to a spot size $(1/e^2)$ $<\,2\,\mu m$ by an achromatic objective (numerical aperture\,=\,0.9) enabling the investigation of single nanoantennas. The results for the fabricated structures are shown in Fig. \ref{fig:fabrication}(e). All plasmonic nanostructures exhibit distinct surface plasmon resonances which peak between 1.4\,eV and 1.5\,eV and can be tuned over a broad spectral range by changing the size of the particles \cite{schraml2014optical,biagioni2009cross,pompa2006metal}. The peak height of the reflectivity spectra are governed by the reflectivity of the substrate and by the ratio between the (geometrical) scattering cross section of the investigated structures compared to the spot size of the excitation laser. Consequently, the triangle array shows the highest value, since the particles are densely packed with an interparticle distance of 350\,nm and, thus, multiple nanotriangles lie within the excitation spot.

   \begin{figure}[b]
   \begin{center}
   \includegraphics[width=\columnwidth]{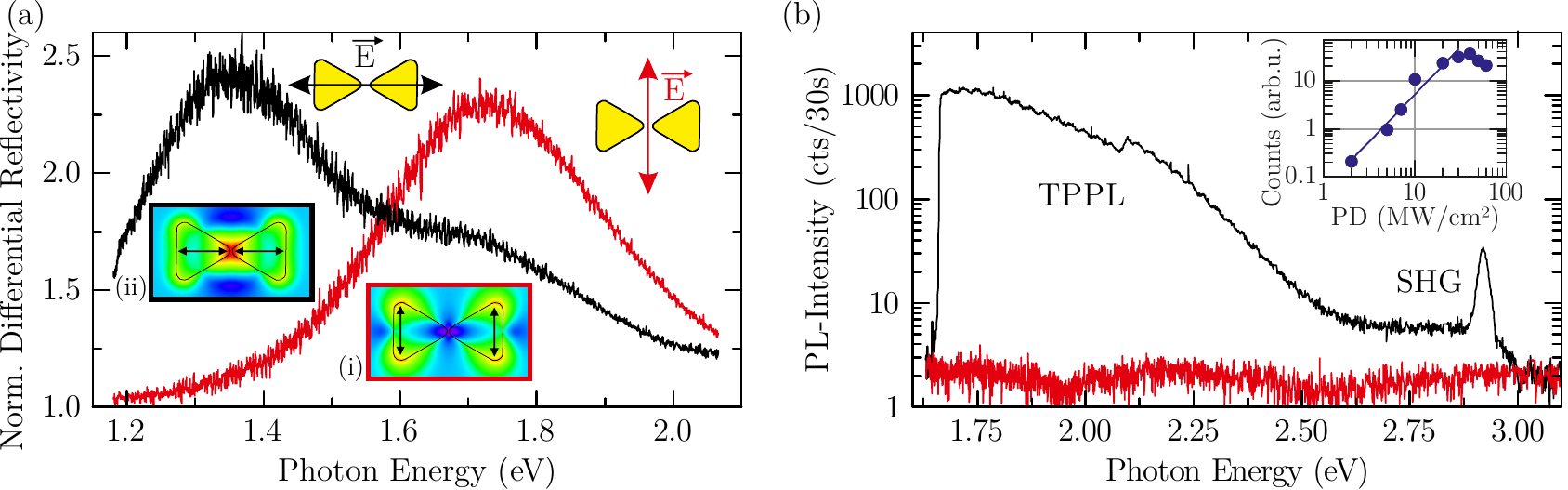}
   \end{center}
   \caption{\label{fig:studies}(a) Measured surface plasmon resonances for incident light polarizations along (black) and perpendicular (red) to the bowtie axis. (i),(ii) Simulated electric field enhancements for both polarizations. (b) Corresponding non-linear photoluminescence intensity. (Inset)  Integrated photoluminescence signal as a function of power density PD. }
  
   \end{figure} 

We continue to present selected results of the optical characterization of a single bowtie ($s=140$\,nm, $g=$10\,nm) on a glass substrate. In Fig. \ref{fig:studies}(a), we show $\gamma$ as a function of photon energy for incident light polarizations perpendicular (red curve) and parallel (black curve) to the bowtie axis. In the first case, we excite an electron oscillation along the outer edges of the bowtie. As shown in the inset (i), there are no overlapping electric near fields in the feed-gap region and consequently no interaction between the triangles takes place\cite{schraml2014optical}. Therefore, we essentially probe the surface plasmon resonance of the individual triangles peaking at $E_{SPR}=1.72$\,eV. If we now turn the polarization by $\mathrm{90^{\circ}}$, the electrons oscillate along the bowtie axis. Both triangles generate strong fields in the feed-gap which overlap and lower the restoring force for the electrons in the neighboring triangle. Consequently, the resonance redshifts by 0.37\,eV to $E_{SPR}=1.35$\,eV\cite{schraml2014optical}.

To probe the non-linear response of the system within the feedgap, we excite our structure with a 100\,fs pulsed titanium-sapphire laser (repetition rate\,=\,82\,MHz) at a photon energy of $E_{ph}=1.46$\,eV and a calculated power density $PD =10\mathrm{\,MW/cm^2}$ at the sample surface. A 1.65\,eV shortpass filter cuts off the excitation light which is polarized along the x-axis of the bowtie. As shown in Fig. \ref{fig:studies}(b), we observe a broad emission at the high energy side of the laser which extends all the way from the filter edge up to the detection limit of the used optics at $\sim 3$\,eV. We attribute this emission to two-photo photoluminescence (TPPL)\cite{PhysRevB.68.115433,biagioni2009dependence,bouhelier2005surface,ko2010nonlinear,schuck2005improving} originating from the tip region of the gold bowtie. In addition, we observe a peak at exactly twice the laser energy $E_{SHG}=2.92$\,eV, which can be attributed to second harmonic generation (SHG) originating from the vicinity of the feedgap\cite{hubert2007role,mcmahon2006second,ko2010nonlinear,hanke2009efficient}. To confirm that the signal arises from a non-linear optical process, we performed power dependent measurements. As shown in the inset of Fig. \ref{fig:studies}(b), the emission intensity scales quadratically with excitation power, which clearly demonstrates the non-linear origin of the observed light. Furthermore, we neither observe any emission at an unpatterned gold film under the same excitation condition nor for a polarization perpendicular to the bowtie axis (red curve). Hence, we conclude that the non-linear effects become only visible due to the presence of the very strong electric field enhancement in our optimized bowtie nanoantennas. Furthermore, the strong dependence of the field enhancement on the incident light's polarizations enables us to switch the non-linear conversion efficiency in a very simple way.

\section{Conclusion} 
\label{sec:conclusion}

In summary, we have optimized the bowtie-geometry theoretically, fabricated structures with high quality and showed enhanced non-linear light conversion experimentally. Our FDTD simulations revealed that the lightning rod effect and the coupling between the triangles composing the bowtie nanoantenna boost the electric field intensity enhancement up to \textgreater 4500 times. It is, therefore, a very well suited geometry to observe linear and non-linear optical effects. Furthermore, we observed a strong dependence of the surface plasmon resonance on the shape of the particle. We chose a thickness of 35\,nm as a good trade-off between high energetic resonances and high field enhancements. Our simulation predicted a reduction of the mean electric field enhancement in the feed-gap from 2300 to 1300 when using a 5\,nm titanium adhesion layer. Using electron beam lithography we realized the simulated structures with precision down to 10\,nm on semiconducting and insulating substrates. We showed that these structures exhibit enhanced two photon photoluminescence and second harmonic generation arsing from the feed-gap of the bowties. Furthermore, the non-linear emission can be controlled by the polarization of the incoming light. The huge electric field enhancement in a bowtie nanoantenna strongly enhances non-linear and other nanophotonic effects at deep subwavelength dimensions.

\appendix    

\acknowledgments     
We acknowledge financial support of the DFG via the SFB 631, Teilprojekt B3 and the German Excellence Initiative via the Nanosystems Initiative Munich. The authors gratefully acknowledge the support of the TUM International Graduate School of Science and Engineering (IGSSE).


\end{document}